\documentclass[12pt]{iopart}
\usepackage{epsfig,cite}
\newcommand{\eq}[1]{\begin{eqnarray} #1 \end{eqnarray}}

\begin{document}

\title[Ratio Fluctuations in HSD Transport Approach]{\ $K/\pi$, $K/p$ and $p/\pi$ Ratio Fluctuations\\within the HSD Transport Approach}

\author{V.~P. Konchakovski $^{1,2}$, M.~Hauer $^1$ M.~I.~Gorenstein $^2$ \\
and E.~L.~Bratkovskaya $^3$}

\address{$^1$ Helmholtz Research School, University of Frankfurt, Germany\\
$^2$ Bogolyubov Institute for Theoretical Physics, Kiev, Ukraine\\
$^3$ Institute for Theoretical Physics, University of Frankfurt, Germany}

\ead{voka@fias.uni-frankfurt.de}

\begin{abstract}
Particle number fluctuations and correlations in nucleus-nucleus collisions at SPS and RHIC energies have been studied within the Hadron-String-Dynamics (HSD) transport approach. Event-by-event fluctuations of pion-to-kaon, proton-to-pion and kaon-to-proton number ratios are calculated for the samples of most central collision events and compared with the available experimental data. It has been found that the HSD model can qualitatively reproduce the measured excitation function for the $K/\pi$ ratio fluctuations in central Au+Au (or Pb+Pb) collisions from low SPS up to top RHIC energies. These predictions impose a challenge for future experiments.
\end{abstract}

\maketitle

\section{Introduction}

The measurement of the fluctuations in the kaon to pion ratio by the NA49 Collaboration was the first event-by-event analysis in nucleus-nucleus collisions \cite{Af01}. It was suggested that this ratio might allow to distinguish events with enhanced strangeness production attributed to the QGP phase \cite{JeKo99}. Nowadays, the excitation function for this observable is available in a wide range of energies:  from the NA49 collaboration \cite{Al08} in Pb+Pb collisions at the CERN SPS and from the STAR collaboration \cite{Da06,Ab09} in Au+Au collisions at RHIC. Results from NA49 show an enhancement of fluctuations in the kaon to pion multiplicity ratio for low energies which may be a signal of a deconfinement phase transition. On the other hand there is no such an enhancement for the proton to pion ratio fluctuations.

This proceeding presents the results of a systematic study of $K/\pi$, $K/p$ and $p/\pi$ ratio fluctuations ($K=K^++K^-$, $\pi=\pi^++\pi^-$, and $p$ means $p+\overline{p}$) based on the HSD transport model \cite{HSD}. For more details we refer readers to~\cite{GoHaKoBr09,KoHaGoBr09}. Event-by-event fluctuations for charged hadron multiplicities also have been studied in the HSD transport approach~\cite{HSD-fluc}.

Let's introduce some notations. The deviation $\Delta N_A$ from the average number $\langle N_A\rangle$ of the particle species $A$ is defined by $N_A=\langle N_A\rangle +\Delta N_A$, while the covariance for species $A$ and $B$ is:
\eq{\label{cov}
\Delta \left(N_A,N_B\right)~\equiv~\langle \Delta N_A
\Delta N_B\rangle~=~\langle N_A N_B\rangle
~-~\langle N_A\rangle \langle N_B\rangle~,
}
the scaled variance and the correlation coefficient
\eq{\label{omega-rho}\fl\hskip2em
\omega_A ~\equiv~\frac{\Delta\left(N_A,N_A\right)}{\langle N_A\rangle} ~=~ \frac{\langle N_A^2\rangle -\langle N_A\rangle^2}{\langle N_A\rangle}~,
~~~~
\rho_{AB}~\equiv~\frac{\langle\Delta N_A~\Delta N_B\rangle}{\left[\langle\left(\Delta N_A\right)^2\rangle ~\langle\left(\Delta N_B\right)^2\rangle\right]^{1/2}}~.
}
The fluctuations of the ratio $R_{AB}\equiv N_A/N_B$ will be characterized by $\sigma^2$ \cite{BaHe99,JeKo99}, which to the second order in $\Delta N_A/\langle N_A \rangle$ and $\Delta N_B/\langle N_B \rangle$ can be written as:
\eq{\label{sigma}
\sigma^2~\equiv~\frac{\langle \left(\Delta R_{AB}\right)^2\rangle}{\langle R_{AB}\rangle^2 }
~\cong~\frac{\omega_A}{\langle N_A\rangle}~+~\frac{\omega_B }{\langle N_B\rangle}
~-~2\rho_{AB}~\left[\frac{\omega_A \omega_B}{\langle N_A\rangle\langle N_B\rangle}\right]^{1/2}.
}

\section{Mixed Events Procedure}

The experimental data for $N_A/N_B$ fluctuations are usually presented in terms of the so called dynamical fluctuations \cite{VoKoRi99}

\eq{\label{sigmadyn}
\sigma_{dyn}~\equiv~\texttt{sign}\left(\sigma^2~-~\sigma^2_{mix}\right)\left|\sigma^2~-~\sigma^2_{mix}\right|^{1/2}~,
}
where $\sigma^2$ is defined by (\ref{sigma}), and $\sigma^2_{mix}$ corresponds to the following mixed events procedure. One takes a large number of nucleus-nucleus collision events, and measures the numbers of $N_A$ and $N_B$ in each event. Then all $A$ and $B$ particles from all events are combined into one {\it set}. A construction of {\it mixed events} is done like the following: One fixes a random number $N=N_A+N_B$ according to the experimental probability distribution $P(N)$, takes randomly $N$ particles ($A$ and/or $B$) from the {\it whole set}, fixes the values of $N_A$ and $N_B$, and returns these $N$ particles into the {\it set}. This is the mixed event number one. Then one constructs event number 2, number 3, etc.

Note that the number of events is much larger than the number of hadrons, $N$, in any single event. Therefore, the probabilities $p_{A}$ and $p_B=1-p_{A}$ to take the $A$ and $B$ species from the whole {\it set} can be considered as constant values during the event construction. Another consequence of a large number of events is the fact that all $A$ and $B$ particles in any constructed {\it mixed event} most probably belong to different {\it physical events} of nucleus-nucleus collisions. Therefore, the correlations between $N_B$ and $N_A$ numbers in a physical event are expected to be destroyed in a mixed event. This is the main purpose of the mixed events construction.

Calculating the $N_A/N_B$ fluctuations for mixed events one gets like in \cite{GoHaKoBr09}:
\eq{ \label{Dmix}
\sigma^2_{mix}~\equiv~\frac{1}{\langle N_A \rangle }~+~\frac{1}{\langle N_B \rangle}~.
}
The mixed event procedure gives the same $\sigma^2$ for $N_A/N_B$ fluctuations as in the GCE formulation for an ideal Boltzmann gas, i.e. $\omega_A=\omega_B=1$ and $\rho_{AB}=0$ in (\ref{sigma}). Note that non-trivial fluctuations of $N_A$ and $N_B$ ($\omega_{A,B}^{mix}\neq 1$) as well as non-zero correlations $\rho_{AB}^{mix}$ may exist in the mixed events procedure, but they are canceled out exactly in $\sigma_{mix}^2$.

\section{HSD Results and Comparison to the Data}

\begin{figure}[ht!]
\centerline{
\epsfig{file=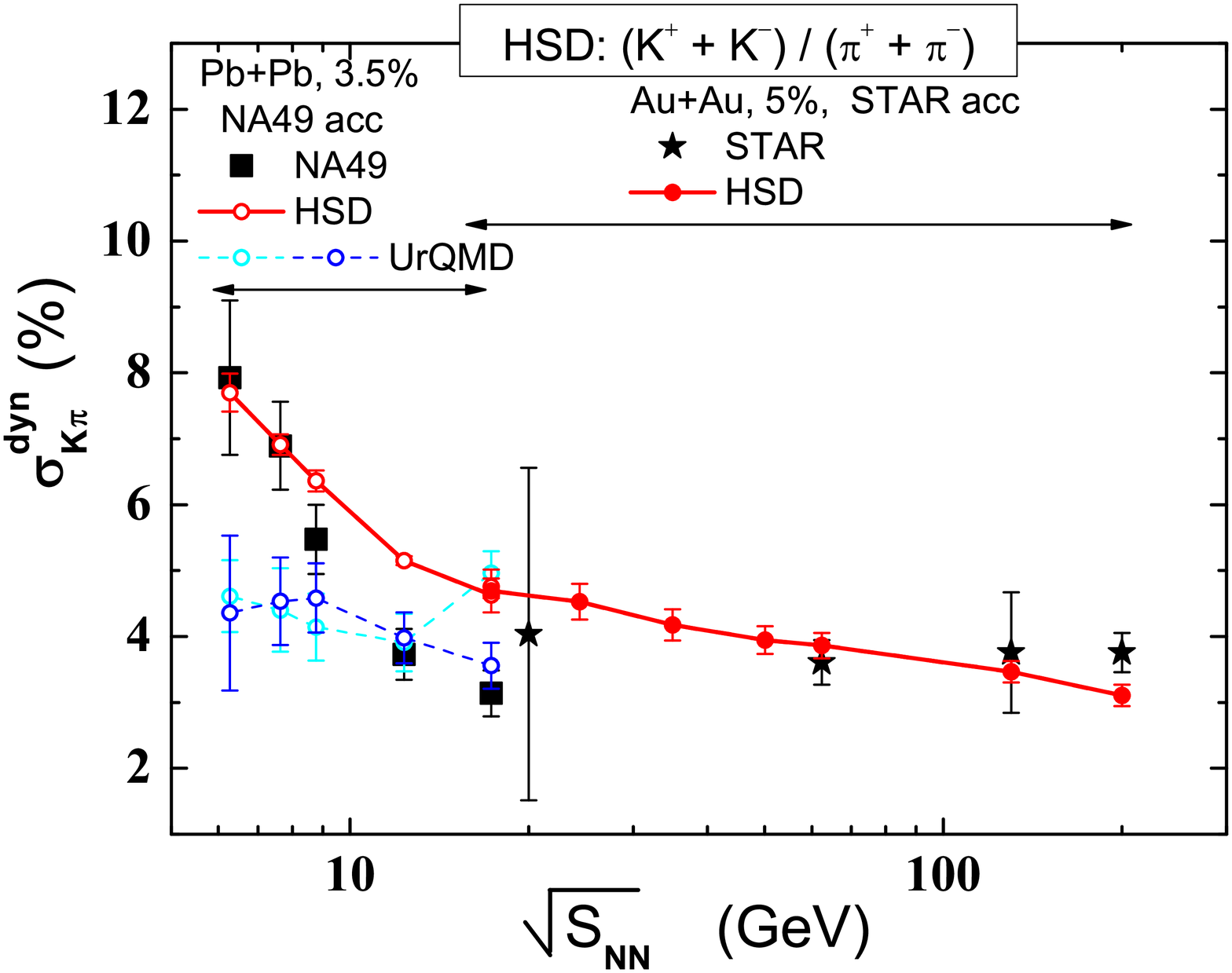,width=0.38\textwidth}\hskip-8pt
\epsfig{file=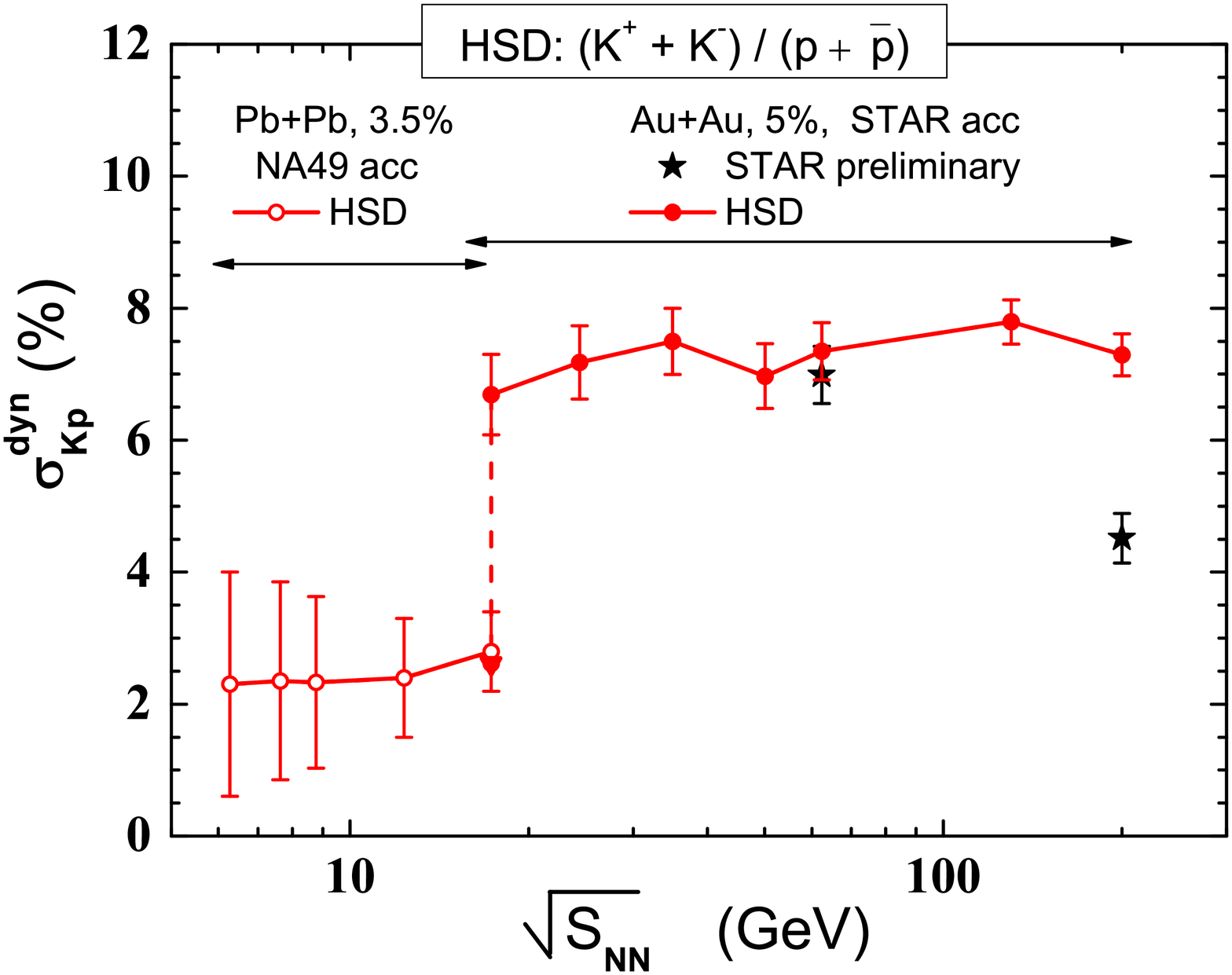 ,width=0.38\textwidth}\hskip-8pt
\epsfig{file=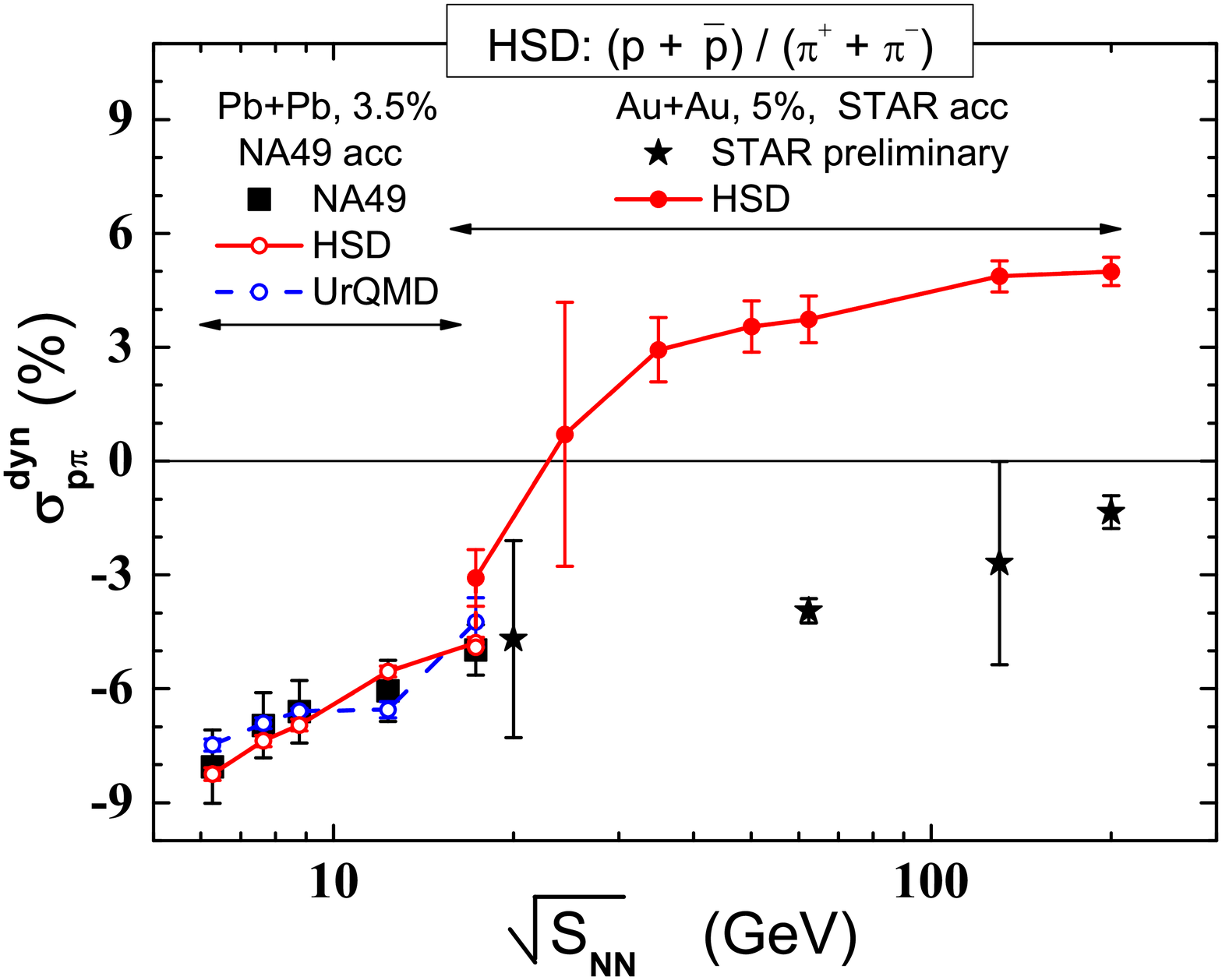,width=0.38\textwidth}}
\caption{
The HSD results for the excitation function in $\sigma_{dyn}$ for the $K/\pi$, $K/p$, $p/\pi$ within the experimental acceptance (solid line) in comparison to the experimental data measured by the NA49 Collaboration at SPS \cite{Al08} and by the STAR Collaboration at RHIC \cite{Da06,Ab09,WeQM09,TiQM09}. The UrQMD calculations are shown by dotted lines.}
\label{fig:fig5}
\end{figure}

The HSD results as well as statistical model calculations for ratio fluctuations in full acceptance are presented in \cite{GoHaKoBr09,KoHaGoBr09} A comparison of the SM results for fluctuations in different ensembles with the data looks problematic at present due to the difficulties in implementing the experimental acceptance and centrality selection which, however, can be taken into account in the transport approach. In order to compare the HSD calculations with the measured data, the experimental cuts are applied for the simulated set of HSD events. In Fig.~\ref{fig:fig5} the HSD results of $\sigma_{dyn}$ for the $K/\pi$, $p/\pi$ and $K/p$ ratios are shown in comparison with the experimental data by the NA49 Collaboration at the SPS~\cite{Al08} and the preliminary data of the STAR Collaboration at RHIC \cite{Da06,Ab09,WeQM09,TiQM09}. The available results of UrQMD calculations (from Refs.  \cite{Al08,Ro04,KrFr06}) are also shown by the dashed lines.

For the SPS energies we use the NA49 acceptance tables from Ref. \cite{Al08}. For the RHIC energies we use the following cuts: in pseudorapidity, $|\eta|<1$, and in transverse momentum, $0.2<p_T<0.6$ GeV/c for kaons and pions and $0.4<p_T<1$ GeV/c for protons \cite{Da06,Ab09,WeQM09,TiQM09}. Note, that HSD results presented in Fig. \ref{fig:fig5} correspond to the centrality selection as in the experiment: the NA49 data correspond to the 3.5\% most central collisions selected via veto calorimeter, whereas in the STAR experiment the 5\% most central events with the highest multiplicities in the pseudorapidity range $|\eta|<0.5$ have been selected.

One sees that the UrQMD model gives practically a constant $\sigma_{dyn}^{K\pi}$, which is by about $40\%$ smaller than the results from HSD at the lowest SPS energy. This difference between the two transport models may be probably attributed to different realizations of the string and resonance dynamics in HSD and UrQMD: in UrQMD the strings decay first to heavy baryonic and mesonic resonances which only later on decay to `light' hadrons such as kaons and pions. In HSD the strings dominantly decay directly to `light' hadrons (from the pseudoscalar meson octet) or the vector mesons $\rho$, $\omega$ and $K^*$ (or the baryon octet and decouplet in case of baryon number $\pm 1$). Such a `non-equilibrated' string dynamics may lead to stronger fluctuations of the $K/\pi$ ratio.

At the SPS energies the HSD simulations lead to negative values of $\sigma_{dyn}$ for the proton to pion ratio. This is in agreement with the NA49 data in Pb+Pb collisions. On the other hand HSD gives large positive values of $\sigma_{dyn}^{p\pi}$ at RHIC energies which strongly overestimate the preliminary STAR data for Au+Au collisions \cite{WeQM09}. For $\sigma_{dyn}^{Kp}$ only preliminary STAR data in Au+Au collisions are available \cite{TiQM09} which demonstrate a qualitative agreement with the HSD results (Fig. \ref{fig:fig5}). The HSD results for $\sigma_{dyn}^{Kp}$ show a weak energy dependence in both SPS and RHIC energy regions. A peculiar feature is, however, a  strong `jump' between the SPS and RHIC values, seen in the middle panel  of Fig. \ref{fig:fig5}, in the HSD calculations which is caused by the  different acceptances in the SPS and RHIC measurements.

The influence of the experimental acceptance is clearly seen at 160 A  GeV where a switch from the NA49 to the STAR acceptance leads to the  jump in $\sigma_{dyn}^{Kp}$ by 3\% - middle panel of Fig.  \ref{fig:fig5}. On the other hand, our calculations for Pb+Pb (3.5\%  central) and for Au+Au (5\% central) collisions - performed within the  NA49 acceptance for  both cases at 160 A GeV - shows a very week  sensitivity of $\sigma_{dyn}^{Kp}$ on the actual choice of the  collision system and centrality -- cf. the coincident open circle and  triangle at 160 A GeV in the middle panel of Fig. \ref{fig:fig5}.

\section{Summary}

The event-by-event multiplicity fluctuations of $K/\pi$, $K/p$ and  $p/\pi$ ratio fluctuations in central Au+Au (or Pb+Pb) collisions from  low SPS up to top RHIC energies have been studied within the HSD  transport approach.

It has been found that the HSD model can qualitatively reproduce the  measured excitation function for the $K/\pi$ ratio fluctuations in  central Au+Au (or Pb+Pb) collisions from low SPS up to top RHIC  energies.  Accounting for the experimental acceptance as well as the  centrality selection has a relatively small influence on $\sigma_{dyn}$  and does not change the shape of the $\sigma_{dyn}$ excitation  function.

The HSD results for $\sigma_{dyn}^{p\pi}$ appear to be close to the  NA49 data at the SPS. The data for $\sigma_{dyn}^{Kp}$ in Pb+Pb  collisions at the SPS energies will be available soon and allow for  further insight.  A comparison of the HSD results with preliminary STAR  data in Au+Au collisions at RHIC energies are not fully conclusive:   $\sigma_{dyn}$ from HSD calculations is approximately in agreement with  data \cite{TiQM09} for the kaon to proton ratio, but overestimate the  experimental results \cite{WeQM09} for the proton to pion ratio. New  data on event-by-event fluctuations in Au+Au at RHIC energies will help  to clarify the situation.

\section*{Acknowledgments}

This work  was supported by the Helmholtz International Center for FAIR within the  framework of the LOEWE program (Landesoffensive zur Entwicklung  Wissenschaftlich-\"Okonomischer Exzellenz) launched by the State of  Hesse.

\end{document}